\documentclass[12pt]{article}
\input{psfig}
  \usepackage{epsfig}                                     

\def\e{\epsilon}

\def\m{\mu}
\def\n{\nu}

\def\r{\rho}
\def\s{\sigma}

\def\bo{{\raise.15ex\hbox{\large$\Box$}}}               
\def\pr{\prod}                                          
\def\face{{\raise.2ex\hbox{$\displaystyle \bigodot$}\mskip-2.2mu \llap {$\ddot
        \smile$}}}                                      

\def\leftrightarrowfill{$\mathsurround=0pt \mathord\leftarrow \mkern-6mu
        \cleaders\hbox{$\mkern-2mu \mathord- \mkern-2mu$}\hfill
        \mkern-6mu \mathord\rightarrow$}       
\def\dvec#1{\vbox{\ialign{##\crcr
        \leftrightarrowfill\crcr\noalign{\kern-1pt\nointerlineskip}
        $\hfil\displaystyle{#1}\hfil$\crcr}}}           

\def\beq{\begin{equation}}
\def\eeq{\end{equation}}

\def\beqx{\begin{displaymath}}
\def\eeqx{\end{displaymath}}

\def\beqa{\begin{eqnarray}}
\def\eeqa{\end{eqnarray}}
\def\NO{\nonumber}

\def\pl#1#2#3{Phys.~Lett.~{\bf B {#1}} (19{#2}) #3}
\def\np#1#2#3{Nucl.~Phys.~{\bf B {#1}} (19{#2}) #3}

\def\pr#1#2#3{Phys.~Rev.~{\bf D {#1}} (19{#2}) #3}

\catcode`@=11
\def\@citex[#1]#2{\if@filesw\immediate\write\@auxout{\string\citation{#2}}\fi
  \def\@citea{}\@cite{\@for\@citeb:=#2\do
    {\@citea\def\@citea{,\penalty\@m}\@ifundefined
      {b@\@citeb}{{\bf ?}\@warning
       {Citation `\@citeb' on page \thepage \space undefined}}%
\hbox{\csname b@\@citeb\endcsname}}}{#1}}

\def\citer{\@ifnextchar [{\@tempswatrue\@citexr}{\@tempswafalse\@citexr[]}}
\def\@citexr[#1]#2{\if@filesw\immediate\write\@auxout{\string\citation{#2}}\fi
  \def\@citea{}\@cite{\@for\@citeb:=#2\do
    {\@citea\def\@citea{--\penalty\@m}\@ifundefined
       {b@\@citeb}{{\bf ?}\@warning
       {Citation `\@citeb' on page \thepage \space undefined}}%
\hbox{\csname b@\@citeb\endcsname}}}{#1}}
\catcode`@=12

\begin{document}
\date{\mbox{ }}

\title{ 
{\normalsize     
DESY 98-048\hfill{\tt ISSN 0418-9833}\\
April 1998\hfill{\tt hep-ph/9804460}}\\[25mm]
\bf TWO-LOOP GAP EQUATIONS\\ FOR THE MAGNETIC MASS\\[8mm]}
\author{F.~Eberlein\footnote{e-mail: {\tt frankeb@x4u2.desy.de}}\\
Deutsches Elektronen-Synchrotron DESY, 22603 Hamburg, Germany}
\maketitle

\thispagestyle{empty}

\begin{abstract}
  \noindent
One-loop gap equations  have recently been used by several authors
to estimate the 
non-perturbative mass gap in a 3-dimensional gauge theory.
I extend the method to two loops and demonstrate, that the 
resulting gap equation has a real and positive solution 
$m \simeq 0.34 g^2$, which is in good agreement with the one-loop 
results and lattice data.
 \end{abstract}

\newpage  
In the  high-temperature phase of the Standard Model one naively expects
a vanishing Higgs vacuum expectation value and  vector
boson mass.
This leads to the well-known breakdown of 
the perturbative expansion
due to severe infrared divergences in the 
magnetic sector of the theory \cite{vier}.
The infrared problem may be cured by
a non-vanishing 
  magnetic mass, which acts as a cutoff regularizing these divergences.
Its  inverse, the magnetic screening length, determines
the size of non-perturbative effects in the symmetric phase
and is closely related to the confinement scale of the effective 3-dimensional theory which describes the high-temperature
limit of the 4-dimensional finite temperature field theory \cite{funf}.
In an apparently massless 3-dimensional
Yang-Mills theory 
 the gauge coupling $g^2$ carries the dimension of mass, thereby
providing a natural mass scale.
A  popular framework for calculating the dynamically generated
 mass in a 3-dimensional 
$ SU(2) $ gauge theory has recently been via gap equations.

Up to now,  several attempts have been made to evaluate the size of 
this mass gap, all of them at the one-loop level \cite{BP,AN,JP1,C}.
In order to check, whether the whole approach is consistent,
it is crucial to extend the method to two loops.
To my knowledge this letter contains the first
treatment of gap equations in next-to-leading order. The corresponding calculation
is rather involved at the technical level.
 
After briefly presenting the idea behind the gap equation approach
and giving an overview of the  one-loop results,
I will give reasons for the necessity of a two-loop calculation, the results of which will be extensively discussed.

\mbox{ }\\
\noindent
{\bf\large General Strategy}\\

All the models  considered so far start form a
massless Yang-Mills action $ S_G $ in 3-dimensional
Euclidean space
 and then add and subtract
 some gauge-invariant
mass term $S_m$, where the subtracted term enters the perturbation theory
at one loop higher than the added term.
This can be formalized by introducing a loop-counting parameter
$ l $  in the following manner: one rescales all the fields by
$ \sqrt{l} $ and calculates 
with the modified action

\beq
S_{eff}  = {1 \over l} \left( S_G (\sqrt{l}W )+S_m (\sqrt{l}W )\right)
- S_m (\sqrt{l}W )
\;
\eeq

\noindent in a formal $ l $-expansion \cite{JP2}.
Perturbative calculations
are no longer done to a fixed order of the gauge coupling $g$,
but as a power series in $l$, resulting in a rearranged or resummed
perturbation series.
Out of the many choices, which are possible for $S_m$, I will concentrate on
 the 
non-linear $\s$-model for reasons specified below.

The gap equation is a self-consistent condition for the vector boson mass.
The goal is to find the particular size of the tree-level mass-term $ m = Cg^2$
leading to a convergent perturbation series.
In other words, the pole of the transverse part of the (Euclidean)
vector boson propagator should remain at $p^2= - m^2$ to any loop-order, i.e.

\beqa
D_T(p^2)  &=& {1\over p^2+m^2-\Pi_T(p^2)}\NO\\[1ex]
 &\sim & {Z\over p^2+m^2} \;\;\;\;\;\;\;\mbox{ for }\; p^2 \sim - m^2
\;,\label{detee}
\eeqa
with some residue $Z$. With eq.~(\ref{detee}) 
one obtains the desired gap equation for the self-energy
in resummed perturbation theory

\beq
\Pi_T (p^2= - m^2) \left( 1+{\partial \Pi_T\over \partial p^2}(p^2=-m^2)\right) = 0
\;.\label{gap}
\eeq
In $n$-th order of resummed perturbation theory one calculates
eq.~(\ref{gap}) up to $l^n$ and solves the gap equation
for $m$. 

At one-loop eq.~(\ref{gap}) reduces to 
\beq
\Pi_T^{1-loop}(p^2= - m^2)  = 0
\;.\label{gap1}
\eeq
In theories with a BRS-symmetry the position of the pole of the propagator
and therefore  eq.~(\ref{gap}) is
 gauge-independent on mass-shell \cite{rebhan}.
The self-energy itself is not gauge-invariant on mass-shell except at one-loop level.

\mbox{ }\\
\noindent
{\bf\large One-Loop Calculations}\\

In recent years, several authors have proposed models to extract a gap mass
at one-loop level.
When studying the electroweak phase transition Buchm\"uller and 
Philipsen could obtain a non-vanishing vector boson mass
in the symmetric phase of a linear 3-dimensional $SU(2)$-Higgs model \cite{BP}.
A mass resummation was supplemented by a vertex resummation
in order to get a BRS-invariant resummed tree-level action resulting in   
a gauge-independent gap equation.
Deeply in the symmetric phase the value for the gap mass
is approximately the same as the one obtained in a
non-linear $\s $-model. This suggests to investigate first 
the two-loop effects in this simpler model.

In order to minimize the amount of diagrams in a two-loop
calculation I follow a suggestion by Jackiw and Pi \cite{JP2}.
The functional integral for the partition function
in the non-linear $\s$-model, where $S_m = S_\s$, reads
\beq
Z = \int DW D\pi \triangle exp -{1\over l}(S_G+S_\s+S_{GF}-lS_\s)
\;,\label{nonlin}
\eeq
with $S_{GF}$ being some gauge-fixing term, which depends only on $W_\m$, and $\triangle$ the corresponding
Fadeev-Popov determinant.
They integrated out the Goldstone and ghost fields 
exactly (in an arbitrary gauge)
and arrived at 
a massive Yang-Mills theory without any additional gauge fixing
terms,
\beq
Z \propto \int DW  exp \left[-{1\over l}\left(
S_G(\sqrt{l}W ) + m^2\mbox{Tr}\int d^3x \sqrt{l}W_\mu \sqrt{l}W_\mu
- l m^2\mbox{Tr}\int d^3x \sqrt{l}W_\mu \sqrt{l}W_\mu \right)\right]
\;.\label{masym}
\eeq
One can also view this massive Yang-Mills theory as a non-linear
$\s$-model (with $R_\xi$-gauge fixing) in  unitary gauge.
A calculation of the self-energy to one-loop in both theories
indeed yields the same result
on mass-shell and therefore the same gap mass.
The off-shell self-energies coincide only in the limit 
$\xi \to \infty$.
In unitary gauge the result for the 
off-shell transverse self-energy
is
\beq
\Pi_T^{1-loop} (p^2) = - {1\over 16\pi} l g^2 m 
\left[ \left( {p^6\over 4m^6}-{2p^4\over m^4}-{10p^2\over m^2}+8\right)
{2m\over p} \mbox{arctan} {p\over 2m}
+{p^4\over m^4}-{4p^2\over m^2} -8\right] 
\;.\label{pistr}
\eeq

The gap equation for the massive Yang-Mills theory to one-loop 
is then a
linear equation for m
and reads
\beq
l m_{SM}^2 - {1\over 16\pi}({63\over 4}\mbox{ln} 3 - 3)\; l g^2 m_{SM} = 0\, ,
\;\label{1gap}
\eeq
or
\beq
m_{SM} \simeq 0.28 g^2
\;.\label{1agap}
\eeq
Note that in unitary gauge the longitudinal part of the self-energy
vanishes for all external momenta, $\Pi_L (p^2) = 0$.

For the 3-dimensional $SU(2)$ gauge theory also other gap equations
 have been considered,
which are based on the Chern-Simons eikonal and on the non-local
action
\beq
S_m^{(JP)} = m^2 \mbox{Tr} \int d^3 x F_{\mu} {1\over D^2} F_{\mu}
\;,\label{jp}
\eeq
where $F_{\mu} = {1\over 2} \e_{\mu\nu\r} F_{\nu\r}$ \cite{AN,JP1}.
Interestingly, the one-loop gap equation of Alexanian and Nair yields
a magnetic mass closely related to $m_{SM}$,
\beq  
m_{AN} = {4\over 3} m_{SM} \simeq 0.38 g^2
\;.\label{an}
\eeq
Jackiw and Pi obtain a complex magnetic mass with the mass term of
eq.~(\ref{jp}),
which, however, can be
modified such that the generated mass gap becomes real.
Another attempt was recently made by Cornwall \cite{C}.
His pinch-technique
gap equation
led to a mass gap of
\beq
m_{C} \simeq 0.25 g^2
\;.\label{PT}
\eeq
It is also very encouraging, that these analytically calculated
gap masses are consistent with the propagator mass 
 obtained in a numerical lattice simulation in Landau gauge \cite{karsch},
\beq
m_{SM}^L = 0.35(1) g^2
\;.\label{L}
\eeq

\mbox{ }\\
\noindent
{\bf\large Two-Loop Gap Equation}\\

\noindent Why is it crucial to perform a two-loop calculation?

\begin{itemize}
\item The loop expansion does not correspond to an  expansion in a small
parameter. Nevertheless, it might very well be,
that the one-loop results provide reasonable approximations
of the true mass gap. This can only be clarified by a two-loop calculation.
\item If the whole method is consistent, the numerical values
for the mass gap in the different models should converge at higher loop-orders,
since to all orders they describe the same Yang-Mills theory.
In this case one expects the two-loop correction to be of order 
$m_{AN}-m_{SM}$.
\item The two-loop gap equation is quadratic in $m$, whereas at one loop it is
linear.
The existence of a positive solution is 
a  non-trivial check of the whole approach.
\end{itemize}

\noindent As mentioned above, we first consider the Lagrangian of  a  
massive YM 
(a resummed non-linear $\sigma$-model in  unitary gauge)

\beq
{\cal L} = {1\over 4} F^a_{\mu\nu} F^a_{\mu\nu} + {1\over 2} m^2 
W^a_{\mu} W^a_{\mu} - {l\over 2} m^2 
W^a_{\mu} W^a_{\mu}
\;,
\eeq
with
\beq
F^a_{\mu\nu} = \partial_{\m}W^a_{\n}-\partial_{\n}W^a_{\m}+\sqrt{l} g 
\e^{abc} W^b_{\m} W^c_{\n},\,\; G=SU(2),\,\; d=3-2\e
\;.
\eeq

\begin{figure}
\begin{center}
\psfig{file=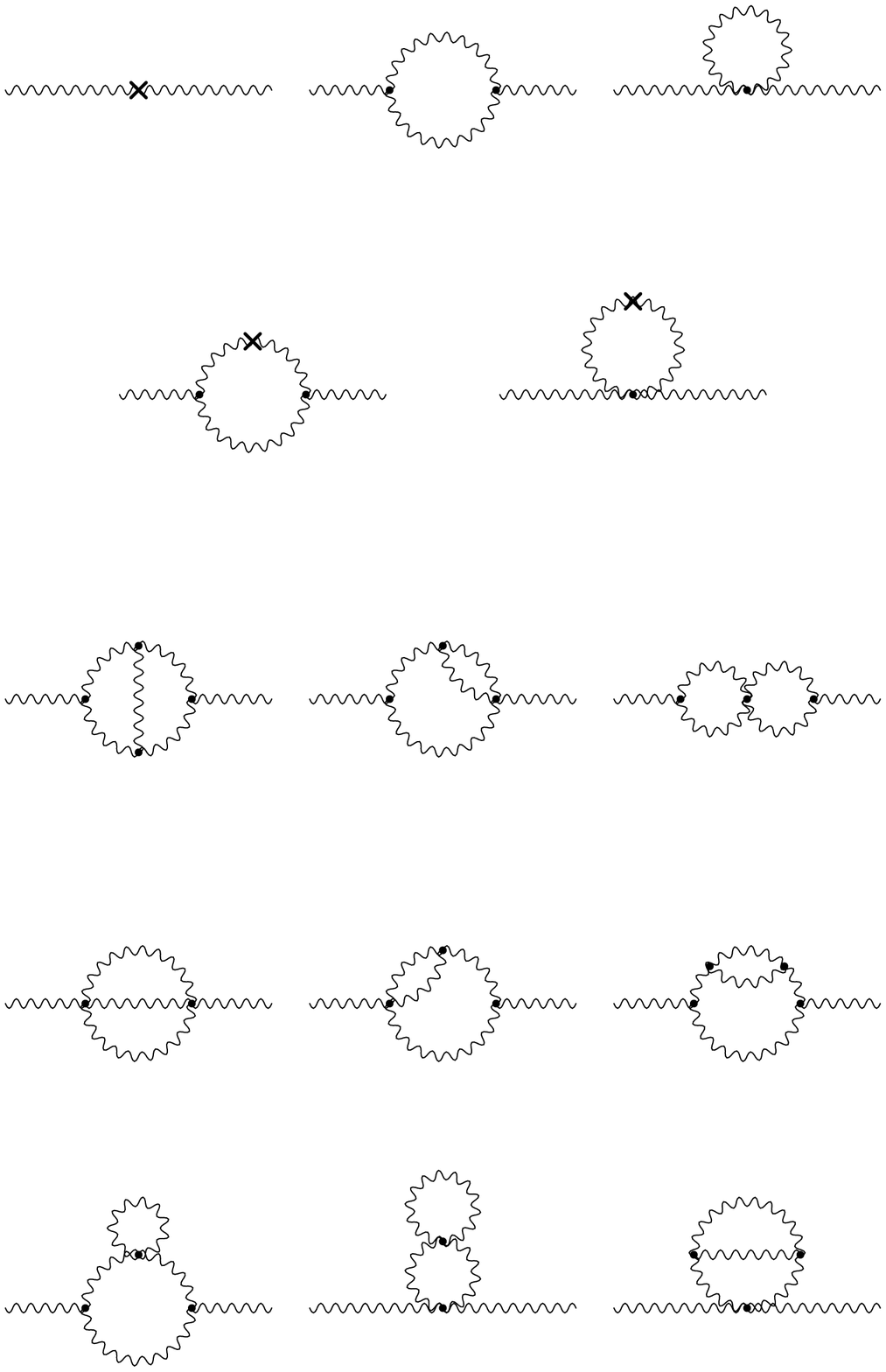}
\caption{ \it Diagrams contributing to the two-loop gap equation in unitary gauge\label{fig}}
\end{center}
\end{figure}

Eq.~(\ref{gap}) has to be expanded up to $O(l^2)$, which requires the evaluation of 
the diagrams 
 depicted in fig.~\ref{fig}.
The contribution of diagrams 1, 2 and 3 to the transverse on-shell
self-energy
was already calculated at one loop,
cf.~eq.~(\ref{1gap}).
The contributions of diagrams 4 and 5 are easily evaluated,  
\beq
\Pi_T^{1-loop-CT} (p^2= - m^2) = {1\over 8\pi}\left( {21\over 4}
\mbox{ln} 3 - 9\right) l^2 g^2 m
\;.\label{ha}
\eeq
 For ${\partial \over \partial p^2}\Pi_T$, which contributes
to ${\cal O}(l {g^2\over m})$, only one-loop 
diagrams are needed in the two-loop gap equation, 
  which can directly be obtained {}from 
eq.~(\ref{pistr}),
\beq
{\partial \over \partial p^2}\Pi_T^{1-loop}(p^2= - m^2) = {1\over 32\pi}\left( 33-{21\over 4}
\mbox{ln} 3 \right) l {g^2\over m}
\;.\label{hi}
\eeq

Far more work has to be done for the evaluation of the remaining
9 two-loop diagrams, which 
contribute
to ${\cal O}(l^2 g^4)$.
As all propagators are massive and the external momentum does not vanish,
the reduction of the scalar integrals to basic integrals with no momenta in the
numerators turns out to be the most difficult step in the calculation.
 For  propagator type integrals this task has been achieved only recently
by Tarasov \cite{Tar}. Using his recurrence relations it is possible 
to reduce the self-energy integrals to a small set of 
linearly independent basic integrals.
For the first time this  method  achieves  a  complete reduction
and stays on an algebraic level as far as possible.
Since the recurrence relations are in some cases quite involved,
they have to be implemented into a FORM package \cite{verm}. In  unitary gauge 
the situation is even more complex due to
the high powers of momenta in the numerator.

In $3-2\epsilon $ dimensions, the  result of the reduction to basic integrals
 for the 
on-shell self-energy  in 
unitary gauge reads

\def\sou#1{\;\parbox{5mm}{\setlength{\unitlength}{0.1mm}
\begin{picture}(50,40)\thicklines#10
\end{picture}}}
\def\sour#1{\;\parbox{10mm}{\setlength{\unitlength}{0.1mm}
\begin{picture}(100,50)\thicklines#1
\end{picture}}}
\def\sourc#1{\;\parbox{15mm}{\setlength{\unitlength}{0.15mm}
\begin{picture}(70,50)\thicklines#1
\end{picture}}}
\def\source#1{\;\parbox{15mm}{\setlength{\unitlength}{0.15mm}
\begin{picture}(70,50)\thicklines#1
\end{picture}}}
\def\eins{\source{\put(0,25){\line(1,0){20}}\put(35,25){\circle{30}}\put(65,25){\circle{30}}\put(80,25){\line(1,0){20}}}}
\def\zwei{\source{\put(0,25){\line(1,0){25}}\put(50,25){\circle{50}}\put(25,0){\oval(50,50)[tr]}\put(75,25){\line(1,0){25}}}}
\def\drei{\source{\put(0,25){\line(1,0){25}}\put(50,25){\circle{50}}\put(50,0){\line(0,1){50}}\put(75,25){\line(1,0){25}}}}
\def\vier{\source{\put(0,25){\line(1,0){55}}\put(27.5,37.5){\circle{25}}\put(67.5,25){\circle{25}}\put(80,25){\line(1,0){20}}}}
\def\funf{\source{\put(0,15){\line(1,0){100}}\put(30,30){\circle{30}}\put(70,30){\circle{30}}}}
\def\sechs{\sourc{\put(0,10){\line(1,0){70}}\put(35,30){\circle{40}}\put(35,10){\line(0,1){40}}}}
\def\sieben{\source{\put(0,25){\line(1,0){100}}\put(50,25){\circle{50}}}}
\def\kdrei{\sour{\put(0,25){\line(1,0){25}}\put(50,25){\circle{50}}\put(50,0){\line(0,1){50}}\put(75,25){\line(1,0){25}}}}
\def\sechs{\sour{\put(0,10){\line(1,0){70}}\put(35,30){\circle{40}}\put(35,10){\line(0,1){40}}}}
\def\sieben{\sour{\put(0,25){\line(1,0){100}}\put(50,25){\circle{50}}}}
\def\ksechs{\sour{\put(10,10){\line(1,0){30}}\put(25,25){\circle{30}}\put(25,10){\line(0,1){30}}}}
\def\ksieben{\sour{\put(0,25){\line(1,0){100}}\put(50,25){\circle{50}}}}

\beqa
{1\over l^2 g^4}\Pi^{2-loop}_T (p^2=-m^2)&=& -{3 \over 128}m^2 \eins  -{1329\over 64}m^2 \zwei \NO\\[1ex]
&& +{849\over 32}m^4\drei +{369\over 16}\vier-{543\over 160 m^2}\funf \NO\\[1ex]
&& +\left({1143\over 64}-33\epsilon\right)\sechs+\left({71917\over 600}\epsilon-{5523\over 320}\right)\sieben  . \label{too}
\end{eqnarray}

Except for $ \kdrei $, which has to be evaluated numerically,
there exist analytic expressions for the basic integrals
in $3-2\epsilon $ dimensions \cite{raj}. The result of
(\ref{too}) was also obtained using a FORM package written independently 
by O.~Tarasov.
Neglecting the resummation counter-terms, which spoil BRS-invariance,
I also calculated the self-energy for the non-linear $\sigma$-model in Feynman gauge, $\xi = 1$. I obtained the same position for the pole of the propagator as in  unitary gauge, which constitutes a very stringent test for the algorithm I used.

Two further remarks have to be made concerning the two-loop calculation. First,
the longitudinal part of the self-energy in  unitary gauge vanishes for all momenta at two-loop level, which is  another  nice
check of the calculation.
Second, the non-linear sigma model is non-renormalizable.
This is not a problem at one loop, since the 3-dimensional self energy is finite in dimensional regularization.
At two loops, however, $\sechs$ and $\sieben$ are UV-divergent,
which requires the addition of  counter-terms ($\sim l^2$) to the Lagrangian. The explicit calculation in Feynman gauge shows that a mass and wave function renormalization
is sufficient to remove the infinities in the self-energy,

\beq
{1\over l^2 g^4}\Pi_{T,\xi = 1}^{2-loop}(p^2) = \left({7\over 12}-{1\over 60} {p^2\over m^2}
\right) {1\over 64 \pi ^2 \e} + \mbox{finite}
\;.
\eeq

Compared to the unitary gauge, where ghosts and Goldstone bosons are
integrated out, the Feynman gauge involves the evaluation of many more 
diagrams, 33 generic two-loop graphs instead of 9.
Note also, that the unitary gauge is not suitable for  renormalization.
Even in renormalizable theories the bad high-energy behavior of the 
propagator
leads to terms $\left({p^2\over m^2}\right)^n {1\over \e}, n>1$, in the self-energy, which
cannot be dealt with by a mass or wave function renormalization.
Renormalization in Feynman gauge introduces a renormalization scale $\m$.
Using the $\overline {MS}$-scheme it turns out that there is almost no 
numerical dependence of the two-loop gap mass on the 
 scale $\m = \mu_{\overline {MS}}$.

Now we are ready to discuss the two-loop gap equation. In an arbitrary gauge of the resummed non-linear $\s$-model it reads

\beqa
&&l m^2 - 0.28455\; l g^2 m + f_1(\xi )\; l^2 g^2 m + f_2(\xi )\; l^2 g^2 m \NO\\[1ex]
&& - \;  0.064346\; l^2 g^4+ \; 0.0037995 \; l^2 g^4\;\mbox{ln}{\m\over  m}  = 0
\;,\label{hap}
\eeqa

\noindent with

\beqa
f_1(\xi ) &=&  {1\over 8\pi}\left( {21\over 4}
\mbox{ln} 3 - 9 +{1\over 4\sqrt{\xi }} \mbox{ln} 3  +\sqrt{\xi } (3- \mbox{ln} 3)\right) \;,\NO\\[1ex]
f_2(\xi ) &=&  {1\over 8\pi}\left({33\over 4} - {21\over 16}
\mbox{ln} 3 +(\xi -{1\over 4}) \mbox{ln} {2\sqrt{\xi}+1\over 2\sqrt{\xi}-1}   -3\sqrt{\xi } \right) 
\;,\; \xi > {1\over 4}\; .
\eeqa

The two-loop gap equation is not exactly gauge parameter independent.
There remains a  weak gauge dependence stemming {}from  diagrams which involve
the resummation counter-terms (mass counter-terms for the vector, ghost and Goldstone field).
In the unitary gauge $f_1$ and $f_2$ reduce to eq.~(\ref{ha}) and (\ref{hi}).
Note that the limit $\xi \to \infty $ has to be taken before divergent integrals
are evaluated \cite{jak}. The gap equations in unitary and Feynman gauge 
turn out to be identical.

\mbox{ }\\  
\noindent
{\bf\large Two-Loop Results}\\

The major result is that the gap equation (\ref{hap}) has indeed a real and positive solution for
${m\over g^2}$. The results are given in table 1 for different values of $\m$ and $\xi$. The two-loop correction to the one-loop gap mass (\ref{1agap}) is only $15 - 20\%$.

\renewcommand{\arraystretch}{1.5}
\setlength{\arrayrulewidth}{0.3mm}

\begin{center}
\begin{tabular}{|c|c|c|c|}

 \hline
${\mu\over m}$ & $0.3$ & $1$ & $3$  \\
\hline
${m\over g^2}, \xi = 1,\infty $ & $0.343$ & $0.335$ & $0.327$ \\
\hline
${m\over g^2}, \xi = 2 $ & $0.345$ & $0.336$ & $0.328$ \\
\hline
${m\over g^2}, \xi = 10 $ & $0.350$ & $0.342$ & $0.334$ \\
\hline  
 \end{tabular}
\end{center}
\begin{center}
 Table 1: {\it Solutions of the two-loop gap equation}
\end{center}

One may worry about the dependence of the gap mass on the renormalization scale $\m$
and on the gauge parameter $\xi$. This is an artefact of (resummed) 
perturbation theory, which is 
 expected to be cancelled at higher orders.
Fortunately, the dependence of ${m\over g^2}$ on $\mu$ and $\xi$ is numerically unimportant. This suggests that the solution  constitutes
 a reliable approximation
to the exact gluon propagator mass in $SU(2)$ gauge theory.

 A vector boson mass $\sim 0.34  g^2$ is not in contradiction with confinement. It is of the same size as the confinement scale given by the string tension which was calculated in \cite{tepi}. The connection of such a propagator mass 
to the heavier
glueball masses  $\sim {\cal O}(1) g^2$  \cite{ote}  in a 3-dimensional $SU(2)$ gauge theory 
  remains to be clarified.

The two-loop result survives all the crucial tests which have been mentioned above unexpectedly well: the quadratic gap equation has a real and positive solution, which is not far away {}from the one-loop result. Moreover, it is now in better agreement with Alexanian and Nair's gap mass and matches perfectly the lattice result obtained by Karsch et.~al.

To judge the significance of a calculation in the non-renormalizable non-linear sigma model, it is crucial to 
perform the whole calculation in the linear Higgs model, which is super-renormalizable.
The linear  model involves the summation of hundreds of two-loop diagrams.
 I have shown that the pole of the Higgs and of the
vector boson propagator
is gauge-invariant to two loops, which is a very  powerful 
test of Tarasov's algorithm and my FORM package.
More importantly, the two-loop gap equation in the linear Higgs model
turned out to be  numerically nearly independent of the Higgs mass.
Therefore, the non-linear sigma model remains a very good approximation, as it already proved
to be the case at one loop. 
Detailed results of this work will be published in a forthcoming paper \cite{fe}.

I thank O.~Tarasov for an independent calculation of eq.~(\ref{too})
as well as for helpful discussions. I am also grateful to W.~Buchm\"uller
and O.~Philipsen for valuable comments and suggestions.


\input{psfig}
\end{document}